\documentclass[12pt]{article}
\usepackage[T2A]{fontenc}
\usepackage[utf8]{inputenc}
\usepackage{setspace,amsmath}
\usepackage{bbold}
\usepackage{amsfonts}
\usepackage{amssymb}
\usepackage{amsthm}
\usepackage{cite}
\usepackage{fontenc}
\usepackage{feynmp-auto}
\usepackage[dvips]{graphicx}
\usepackage[unicode,linktocpage=true,plainpages=false,pdfpagelabels=false]{hyperref}
\usepackage{bm}
\usepackage{float}
\newtheorem{theorem}{Theorem}[section]

\newtheorem{lemma}[theorem]{Lemma}

\usepackage{authblk}
 \usepackage{lipsum}

\begin{document}
\title{Quantum equation of motion and two-loop cutoff renormalization for $\phi^3$ model}
\author[1]{A.~V.~Ivanov\thanks{E-mail: regul1@mail.ru}}
\author[2]{N.~V.~Kharuk\thanks{E-mail: natakharuk@mail.ru}}
\affil[1]{\it{St. Petersburg Department of Steklov Mathematical Institute of Russian Academy of Sciences, 27 Fontanka, St. Petersburg, Russia}}
\affil[2]{\it{ITMO University, St. Petersburg 197101, Russia}}
\date{\vskip 5mm}
\maketitle
\vskip 10mm
\begin{abstract}
We present two-loop renormalization of $\phi^3$-model effective action by using the background field method and cutoff momentum regularization. In this paper, we also study a derivation of the quantum equation of motion and its application to the renormalization.
\end{abstract}
\vspace{1cm}


\textit{Acknowledgements:}
This work was supported by the Russian Science Foundation (project 19-11-00131).
A. V. Ivanov is a Young Russian Mathematics award winner and would like to thank
its sponsors and jury.

\normalsize
\newpage

\section{Introduction} 
Renormalization theory (see \cite{111,321,322,323}) plays a crucial role in the quantum field theory and largely depends on a regularization. This work is devoted to a cutoff momentum one, which has its pros and cons. On the one hand, it can break Lorentz and gauge invariance and can add non-logarithmic divergences, but on the other hand, it is a more physical procedure and it preserves dimension. As a rule, to study the properties of regularization and renormalization, we often choose the simplest theory (not necessarily physical), which clearly shows the main process. We are going to work with a scalar $\phi^3$-model, which was used to study  dimensional regularization in the four- (see \cite{2}) and six-dimensional (see \cite{1}) cases as well as for more complex versions of the theory \cite{122,13,14,15,16,17,18}.

In the paper, we study a two-loop renormalization of the scalar $\phi^3$-theory with a cutoff momentum regularization in four and five dimensions (super-renormalizable cases), and in six dimensions (renormalizable case). We use the background field method (see \cite{1022,1032,3,4,5,6}), obtain a quantum equation of motion, and explain its applications to the renormalization process.

First, we need to introduce the Lagrangian density of the Euclidean $\phi^3$-model
\begin{equation}
\label{L}
\mathcal{L}[\phi](x)=\frac{1}{2}\partial_{\mu}\phi(x)\partial^{\mu}\phi(x)+\frac{1}{2}m^2\phi^2(x)-\frac{g}{6}\phi^3(x),  \,\,\, x\in\mathbb{R}^n,
\end{equation}
where $m>0$ is a mass parameter, $g$ is a coupling constant, and $n$ is the dimension. 

Then we can define an action of the theory as 
$S[\phi]=\int_{\mathbb{R}^n}d^nx\,\mathcal{L}[\phi](x)$. Next we assume that the scalar field $\phi$ decreases at infinity; therefore, one can integrate by parts and obtain the crucial property
\begin{equation}
\label{S}
S[\phi+B]=S[B]+(M,\phi)+\frac{1}{2}(N\phi,\phi)-\frac{g}{6}\int_{\mathbb{R}^n}d^nx\,\phi^3(x),
\end{equation}
where the Laplace-type operator $N$ and the field $M$ in the point $x\in\mathbb{R}^n$ are defined by formulae
\begin{equation}
\label{NM}
N(x)=-\partial_{\mu}\partial^{\mu}+m^2-gB(x),\,\,\,\,\,\,M(x)=-\partial_{\mu}\partial^{\mu}B(x)+m^2B(x)-\frac{g}{2}B^2(x),
\end{equation}
and where $B$ is a background field, which is defined below (see Section \ref{quant}).

\section{Problem statement}
\subsection{Green's function and heat kernel} 
Let us introduce some extra definitions related to the operator $N(x)$. We denote by $G(x,y)$ and $K(x,y;\tau)$, respectively, the Green's function and the heat kernel, see \cite{7,8,vas1,vas2}, which satisfy the problems 
\begin{equation}
\label{pr}
N(x)G(x,y)=\delta(x,y),\,\,\,\,\,\,
 \begin{cases}
\left(\frac{\partial}{\partial\tau}+N(x)\right)K(x,y;\tau)=0;\\
K(x,y;0)=\delta(x-y),
 \end{cases}
\end{equation}
for all $x,y\in\mathbb{R}^n$ and $\tau\in\mathbb{R}_+$. Under the conditions described above, we have
\begin{equation}
\label{l1}
\frac{\delta}{\delta B(z)}G(x,y)=g\,G(x,z)G(z,y),\,\,\,\,\,\,
\frac{\delta}{\delta B(z)}K(x,y;\tau)=g\int_0^{\tau}ds\,K(x,z;\tau-s)K(z,y;s).
\end{equation}
To prove the last formulae we need to apply the functional derivative, which satisfies the equality
\begin{equation}
\frac{\delta B(y)}{\delta B(x)}=\delta(x-y),
\end{equation}
to the problems (\ref{pr}) for the Green's function and the heat kernel.

Then we introduce the logarithm of the determinant of the operator $N$ as the following integral (see \cite{vas1,vas2})
\begin{equation}
\ln\det(N/N|_{B=0})=-\int_{\mathbb{R}^n}d^nx\int_{\mathbb{R}_+}\frac{d\tau}{\tau}\,[K(x,x;\tau)-
(4\pi\tau)^{-n/2}e^{-m^2\tau}],
\end{equation}
where we assume that some type of regularization has been used.

Therefore, using the equality for the heat kernel 
\begin{equation}
\int_{\mathbb{R}^n}d^nx\,K(y,x;\tau)K(x,z;s)=K(x,y;\tau+s),
\end{equation}
one can find the first variation in the form
\begin{equation}
\label{ln}
\frac{\delta}{\delta B(x)}\ln\det(N)=-g\,G(x,x).
\end{equation}
This equality makes sense for the regularized objects. Additional properties one can find in  Appendix A.

\subsection{Diagram technique} For clarity, it is convenient to introduce a diagram technique. We  notate the Green's function $G(x,y)$ by a line with two indices $x$ and $y$, and the integral
$\int_{\mathbb{R}^n}d^nx\,\left(\frac{\delta}{\delta\eta(x)}\right)^3$ by a vertex with three external lines, see \cite{pes}. Let us give some examples of using the technique.

\begin{figure}[H]
\centerline{\includegraphics[width=0.52\linewidth]{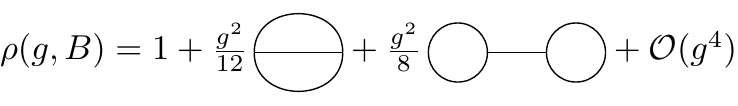}}
\caption{The main terms of the functional $\rho(g,B)$.}
\label{fig1}
\end{figure}

1) Let a functional $\rho(g,B)$ be equal to 
\begin{equation}
\left.e^{\frac{g}{6}\int_{\mathbb{R}^n}d^nx\,\left(\frac{\delta}{\delta\eta(x)}\right)^3}e^{\frac{1}{2}(G\eta,\eta)}\right|_{\eta=0}. 
\end{equation}
It is just a sum of connected vacuum diagrams (and their products). In Fig. \ref{fig1} one can see the first terms of the expansion in powers of the coupling constant $g$. The next correction is multiplied by  $g^4$.

\begin{figure}[H]
\centerline{\includegraphics[width=0.6\linewidth]{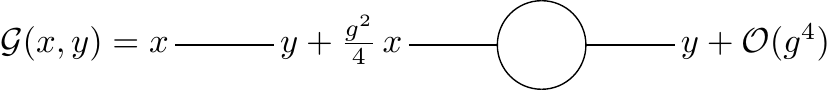}}
\caption{The extended Green's function with first correction.}
\label{fig2}
\end{figure}

2) Let us define an extended Green's function $\mathcal{G}(x,y)$ as a sum of such contributions to the functional
\begin{equation}\label{Green}
\left.\frac{\delta}{\delta\eta(x)}\frac{\delta}{\delta\eta(y)}
e^{\frac{g}{6}\int_{\mathbb{R}^n}d^nx\,\left(\frac{\delta}{\delta\eta(x)}\right)^3}e^{\frac{1}{2}(G\eta,\eta)}\right|_{\eta=0},
\end{equation}
which would become one-particle irreducible (1PI) if the free ends $x$ and $y$ were connected. Then, it takes the form depicted in Fig. \ref{fig2}.
\begin{lemma}\label{Lbi}
Under the conditions described above, the functional (\ref{Green}) contains the function 
$\mathcal{G}(x,y)$ with the coefficient $\rho(g,B)$.
\end{lemma}
This statement can be proved using combinatorial methods and binomial coefficients \cite{cvitanovic-1983,dop2}.

\subsection{Background field method} Primarily, we need to introduce an effective action $W$ as the following path integral, see \cite{1022,1032,3,4,5,6},
\begin{equation}\label{path_int}
e^{-W}=\int_{H}\mathcal{D}\phi\,e^{-S[\phi]},
\end{equation}
where $H$ is a functional set, which is determined using physical reasons. Actually, the effective action is a function of  $H$. Then, according to the background field method, we make the shift $\phi\to\phi+B$. Thus, using formula (\ref{S}), we get
\begin{equation}
\label{b1}
e^{-W[B]}=e^{-S[B]}\int_{H_0}\mathcal{D}\phi\,e^{-(M,\phi)-\frac{1}{2}(N\phi,\phi)+\frac{g}{6}\int_{\mathbb{R}^n}d^nx\,\phi^3(x)},
\end{equation}
where $H_0=\{\phi-B:\phi\in H\}$ is a new set of integration after the shift $H\to H_0$. We suppose that the dependence of $W=W[B]$ on $H$ is dictated by the background field $B$, which is defined below by using the quantum equation of motion. Then we make one more shift $\phi\to\phi+G\eta$, where $G$ is an integral operator with the kernel $G(x,y)$, and $\eta$ is a smooth auxiliary field. In this case, we have
\begin{equation}
\label{b2}
\int_{H_0}\mathcal{D}\phi\,e^{-(M,\phi)-\frac{1}{2}(N\phi,\phi)+\frac{g}{6}\int_{\mathbb{R}^n}d^nx\,\phi^3(x)}=
\det(N)^{-1/2}
\left.e^{-(M,\frac{\delta}{\delta\eta})+\frac{g}{6}\int_{\mathbb{R}^n}d^nx\,\left(\frac{\delta}{\delta\eta(x)}\right)^3}
e^{\frac{1}{2}(G\eta,\eta)}
\right|_{\eta=0},
\end{equation}
where we fixed the normalization property of the measure from formula (\ref{path_int}) by using the condition 
\begin{equation}
\int_{H_0}\mathcal{D}\phi\,e^{-\frac{1}{2}(N\phi,\phi)}=[\det(N)]^{-1/2}.
\end{equation}

\subsection{Quantum equation of motion} 
\label{quant}
Let us obtain the equation of motion. For this purpose, we need to find two kinds of contributions to the effective action $W[B]$. Note that it is possible to write the following decomposition $W[B]=\sum_{n=0}^{+\infty} W_n[B]$, where $W_n[B]$ contains the $M$-vertex $n$ times, see formula (\ref{b2}).
\begin{lemma}
\label{L1} Under the conditions described above a coefficient for $(GM,M)$ in $W_2[B]$, consisting of connected diagrams and their products, is equal to $\frac{1}{2}\rho(g,B)$.
\end{lemma}
\begin{lemma} 
\label{L2}
Under the conditions described above, we have $$W_1[B]=-\frac{g}{2}\rho(g,B)\int_{\mathbb{R}^n}d^nx\,GM(x)\mathcal{G}(x,x).$$
\end{lemma}
\noindent\textbf{Proof:} To find the contribution we need to consider the chain of equalities. The first one is
\begin{equation}
-\left.(M,\frac{\delta}{\delta\eta})e^{\frac{g}{6}\int_{\mathbb{R}^n}d^nx\,\left(\frac{\delta}{\delta\eta(x)}\right)^3}e^{\frac{1}{2}(G\eta,\eta)}\right|_{\eta=0}=
-\left.e^{\frac{g}{6}\int_{\mathbb{R}^n}d^nx\,\left(\frac{\delta}{\delta\eta(x)}\right)^3}(GM,\eta)e^{\frac{1}{2}(G\eta,\eta)}\right|_{\eta=0}.
\end{equation}
Then, we need to  use properties of the functional derivative in the form
\begin{equation}
\left[e^{\frac{g}{6}\int_{\mathbb{R}^n}d^nx\,\left(\frac{\delta}{\delta\eta(x)}\right)^3},(GM,\eta)\right]=\frac{g}{2}
e^{\frac{g}{6}\int_{\mathbb{R}^n}d^nx\,\left(\frac{\delta}{\delta\eta(x)}\right)^3}\left(GM,\frac{\delta^2}{\delta\eta^2}\right).
\end{equation}
Finally, the statement follows from Lemma \ref{Lbi}.
$\blacksquare$

Thus, we can give a definition of the quantum equation of motion. Using Lemmas \ref{L1} and \ref{L2} and varying the combination $\frac{1}{2}\rho(g,B)(GM,M)+W_1[B]$ by the $M$-vertex, one can write down the equation in the form
\begin{equation}
\label{eqmot}
M(x)=\frac{g}{2}\,\mathcal{G}(x,x),
\end{equation}
where $x\in\mathbb{R}^n$. Of course, it contains the divergencies, so we should consider it with the use of some type of regularization. It is  easy to see that the equation is nonlinear with respect to the background field. In a particular case, after regularization, we can express a trace part of the Green's function
\begin{equation}\label{em}
G(x,x)=\frac{2}{g}M(x)+O(g^2).
\end{equation}

Now we can define the background field $B$  as a solution of the problem which consists of the quantum equation of motion (\ref{eqmot}) and an asymptotic behaviour at infinity. The last condition is taken from the definition of $H$.
\begin{theorem}
\label{Teqmot}Under the conditions described above, for all $x \in \mathbb{R}^n$ we have 
\begin{equation}
\frac{\delta}{\delta B(x)}W[B]=M(x)-\frac{g}{2}\,\mathcal{G}(x,x),
\end{equation}
where some type of regularization has been applied.
\end{theorem}

\begin{figure}[H]
\centerline{\includegraphics[width=0.54\linewidth]{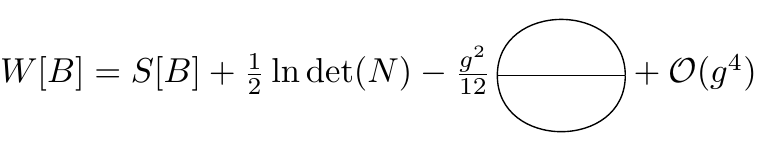}}
\caption{The effective action with the first and the second 1PI corrections.}
\label{fig3}
\end{figure}

The last expression follows from  formulae (\ref{NM}) and (\ref{ln}), and definition of the function $\mathcal{G}(x,y)$.
From equalities (\ref{b1}) and (\ref{b2}) one can express the effective action, which after using Theorem \ref{Teqmot}, has the  form depicted in Fig. \ref{fig3}.

In particular, this means that diagrams such as "glasses" are cancelled.

\subsection{Regularization}
There are many ways to do the regularization (dimensional, Pauli–Villars type, and others). We are going to use the cutoff momentum regularization in a special form. It should be noted that we are interested in infrared divergencies in the coordinate representation. This means one should regularize the Green's function expansion when $x\sim y$. The rules are as follows:
\begin{enumerate}
\item The factor $r^{-k}$ with $k\in\mathbb{N}$ tends to $\chi_{r\Lambda>1}r^{-k}$;
\item The factor $\ln r$ tends to $\chi_{r\Lambda>1}\ln r-\chi_{r\Lambda\leqslant1}\ln\Lambda$,
\end{enumerate}
where $\chi_{(a,b)}$ is a characteristic function of $(a,b)$, and $\Lambda$ is a parameter of the regularization. 
This means that $G^{\Lambda}\to G$ as $\Lambda\to+\infty$ in the sense of generalized functions.
In this case one can write down the trace parts of the Green's function for $n=3, 4, 5, 6$  dimensional cases:
\begin{equation}
\label{gr3}
G_3^{\Lambda}(y,y)=PS_3(y,y);
\end{equation}
\begin{equation}
\label{gr4}
G_4^{\Lambda}(y,y)=\frac{L}{8\pi^2}a_1(y,y)+PS_4(y,y);
\end{equation}
\begin{equation}
\label{gr5}
G_5^{\Lambda}(y,y)=PS_5(y,y);
\end{equation}
\begin{equation}
\label{gr6}
G_6^{\Lambda}(y,y)=\frac{ L}{32\pi^3}a_2(y,y)+PS_6(y,y),
\end{equation}
where the subscript  corresponds to the dimension of the space and $L=\ln (\Lambda/\mu)$.
The last equalities do not violate the limit transition for the Green's function $G^{\Lambda}(x,y)$, they just redefine the value on the diagonal $x=y$.
Of course, after the cutoff regularization is applied, the Green's function has logarithmic $L$ and non-logarithmic $\Lambda$ singularities. The second kind of them has a different nature, so it may not be considered (see, e.g., \cite{Har, Hag}).

\section{Renormalization}
The   renormalization process is based on redefining of the model parameters $m^2$, $\phi$, and $g$. We are going to consider renormalizable case, when $n=6$, and then super-renormalizable cases, when $n=3,4,5$. For the convenience we introduce some extra types for the sign "$=$". The notation IR ($\stackrel{\mathrm{IR}}{=} $) means that both sides of an equality contain the same infrared singular contributions without consideration of parts, proportional to the zero or the first degree of the background field $B$. Note also that we use the logic and notations proposed in \cite{DIF}.

\subsection{n=6 dimensional case}
\label{r6}

In the renormalizable case we have an infinite number of divergencies. Thus we need to find the renormalization constants $Z$, $Z_0$, and $Z_m$. Using the fact that the process of renormalization is equivalent to the transitions
\begin{equation}
\phi\to\sqrt{Z}\phi,\,\,\,\,\,\,g\to Z_0Z^{-\frac{3}{2}}g,\,\,\,\,\,\,m^2\to Z_mZ^{-1}m^2,
\end{equation}
which cancel the singularities, we plan to consider a two-loop renormalization. Using the Lagrange density from (\ref{L}), one can conclude that the only finite number of the coefficients  should be found:
\begin{equation}\label{Z0}
Z_0(g)=1-a_{12}g^2 L-a_{14}g^4 L-a_{24}g^4 L^2+o(g^4);
\end{equation}
\begin{equation}\label{Zm}
Z_m(g)=1-b_{12}g^2 L-b_{14}g^4 L-b_{24}g^4 L^2+o(g^4);
\end{equation}
\begin{equation}\label{Zg}
Z(g)=1-c_{12}g^2 L-c_{14}g^4 L-c_{24}g^4 L^2+o(g^4).
\end{equation}

First, we find the coefficients proportional to $g^2 L$. For this purpose, we need to consider the singularity from the one-loop correction. From  formulae (\ref{ln}) and (\ref{Gd6}), it follows that the singular logarithmic part has the form
\begin{equation}
\ln\det(N)\stackrel{\mathrm{IR}}{=}\frac{ L}{32\pi^3}\int_{\mathbb{R}^6}d^6x\,a_3(x,x).
\end{equation}
Thereby, the contribution to the effective action, see Fig. \ref{fig3}, has the form
\begin{equation}
\frac{1}{2}\ln\det (N)\stackrel{\mathrm{IR}}{=} \frac{g^2 L}{6(4\pi)^3}\frac{(\partial_\mu B,\partial_\mu B)}{2}+
\frac{m^2g^2 L}{(4\pi)^3} \frac{(B,B)}{2}-
\frac{g^3 L}{(4\pi)^3}\frac{(B^2,B)}{6},
\end{equation}
and the coefficients are
\begin{equation}\label{cor1}
c_{12}=\frac{1}{6(4\pi)^3},\,\,\,\,\,\,b_{12}=\frac{1}{(4\pi)^3},\,\,\,\,\,\,a_{12}=\frac{1}{(4\pi)^3}.
\end{equation}

Let us find the coefficients proportional to $g^4 L$. They appear from the two-loop correction. Summing up all the terms from formulae (\ref{6a4})--(\ref{6a8}), and  using the equalities (\ref{6eq}), (\ref{b1}), and (\ref{b2}), we obtain the contribution to the effective action as
\begin{equation}
-\frac{11g^4 L}{36(4\pi)^6}\frac{(\partial_\mu B,\partial_\mu B)}{2}+\frac{m^2g^4 L}{6(4\pi)^6}\frac{(B,B)}{2}-\frac{g^5 L}{6(4\pi)^6}\frac{(B^2,B)}{6}.
\end{equation}
This means that the coefficients are
\begin{equation}
c_{14}=-\frac{11}{36(4\pi)^6},\,\,\,\,\,\,b_{14}=\frac{1}{6(4\pi)^6},\,\,\,\,\,\,a_{14}=\frac{1}{6(4\pi)^6}.
\end{equation}

In the same way, using the formulae (\ref{6a10})--(\ref{6a12}), the contribution proportional to  $g^4 L^2$ is given by
\begin{equation}\label{2loop}
\frac{5g^4 L^2}{36(4\pi)^6}\frac{(\partial_\mu B,\partial_\mu B)}{2}+\frac{5m^2g^4 L^2}{4(4\pi)^6}\frac{(B,B)}{2}-\frac{5g^5 L^2}{4(4\pi)^6}\frac{(B^2,B)}{6}.
\end{equation}

At the same time the one-loop corrections give a contribution to this type of singularity as well. Let us introduce the regularized Laplace operator
\begin{equation}
N^{ren}(x)=-Z\partial_\mu\partial^\mu +Z_mm^2-Z_0gB(x),
\end{equation}
thus we can obtain the following representation
\begin{align}
N^{ren}(x)/Z&=-\partial_\mu\partial^\mu+\frac{Z_m}{Z}m^2-\frac{Z_0}{Z}gB(x)\\
&=N(x)\bigg|_{\substack{m^2\to (Z_m/Z)m^2\\g\to (Z_0/Z)g}\,\,\,\,\,\,}.
\end{align}

Then we get
\begin{equation}
\int_{H_0}\mathcal{D}\phi e^{-\frac{1}{2}(N^{ren}\phi,\phi)}=C(Z)\left[\det (N^{ren}/Z)\right]^{-1/2},
\end{equation}
where $C(Z)$ is a constant, in which we are not interested in.
Therefore, we obtain
\begin{equation}
\frac{1}{2}\ln \det(N^{ren})=-\left.\frac{L}{(4\pi)^3}\int_{\mathbb{R}^6} d^6x\, a_3(x,x)\right|_{\substack{m^2\to (Z_m/Z)m^2\\g\to (Z_0/Z)g}\,\,\,\,\,\,}.
\end{equation}

Substituting  formula (\ref{SD3}) for $a_3(x,x)$, we have the following additional contribution from one-loop correction
\begin{equation}\label{2_vkl2}
\frac{Lg^2}{6(4\pi)^3}\frac{Z_0^2}{Z^2}\frac{(\partial_\mu B,\partial_\mu B)}{2}+\frac{g^2m^2L}{(4\pi)^3}\frac{Z_mZ_0^2}{Z^3}\frac{(B,B)}{2}-\frac{g^3L}{(4\pi)^3}\frac{Z_0^3}{Z^3}\frac{(B^2,B)}{6}.
\end{equation}

Using equations (\ref{Z0})--(\ref{Zg}) and values (\ref{cor1}) from one-loop calculation, we have 
\begin{equation}\label{2_dopv}
-\frac{5g^4L^2}{18(4\pi)^6}\frac{(\partial_\mu B,\partial_\mu B)}{2}-\frac{5g^4m^2L^2}{2(4\pi)^6}\frac{(B,B)}{2}+\frac{5g^5L^2}{2(4\pi)^6}\frac{(B^2,B)}{6}.
\end{equation}

Thus, summing up the contribution from the two-loop divergences (\ref{2loop}) and the contribution from the first loop (\ref{2_dopv}),
we obtain the following values of the coefficients near $g^4L^2$
\begin{equation}
c_{24}=-\frac{5}{36(4\pi)^6},\,\,\,\,\,\,b_{24}=-\frac{5}{4(4\pi)^6},\,\,\,\,\,\,a_{24}=-\frac{5}{4(4\pi)^6}.
\end{equation}

The coefficients, obtained above, are in full agreement with the results obtained earlier (see \cite{1}) in the case of the dimensional regularization. We deliberately disregarded contributions of type (\ref{6a6}). The sum of all such terms equals $-\frac{5}{6}\frac{g^2L}{2(4\pi)^3}\int d^6x\,v(x)PS_6(x,x)$; we consider it in Remark 1 of Sec. \ref{r4}. It should also be noted that the two-loop correction contains a term of the form (see formulae (\ref{6a3}) and (\ref{6a9}))
\begin{equation}
\frac{g\Lambda^2}{2(4\pi)^3}\int_{\mathbb{R}^6}d^6x\,\frac{\delta}{\delta B(x)}\ln\det(N)+\frac{g^2\Lambda^2}{2(4\pi)^6}\int_{\mathbb{R}^6}d^6x\,a_2(x,x).
\end{equation}

It seems that the first term contains a high degree of the field $B$, but it does not. One can use the expansion of the quantum equation of motion in the form (\ref{em}). Therefore, we get
\begin{equation}\label{gB}
\frac{\delta}{\delta B(x)}\ln\det(N)\sim -gB^2(x)+\ldots,
\end{equation}
where the terms proportional to $B^1$, $B^0$, and $\mathcal{O}(g^3)$ are not taken into account. Further, using formula (\ref{SD2}), we can rewrite the contribution as 
\begin{equation}
-\frac{g^2\Lambda^2}{(4\pi)^3}\left(1-\frac{g^2}{2(4\pi)^3}\right)\frac{(B,B)}{2}.
\end{equation}
Actually, the singularity $\Lambda^2$ has a different nature and can be eliminated by  redefining a regularized trace part of the Green's function, or by renormalization of the mass parameter.

\subsection{n=5 dimensional case}
\label{r5}
In the five-dimensional case we have only a finite number of divergencies. From  formula (\ref{gr5}) it follows that the one-loop correction does not have singularities. Thereby, from  equations (\ref{5a1})--(\ref{5a5}) we obtain the contribution to the effective action
\begin{equation}
\label{55}
-\frac{g^4 L}{12(4\pi)^4}\frac{(B,B)}{2}+\frac{g\,\Lambda}{6(4\pi)^2}\int_{\mathbb{R}^5}d^5x\,\frac{\delta}{\delta B(x)}\ln\det(N),
\end{equation}
where the formulae $S^4=\frac{8}{3}\pi^2$ and (\ref{ln}) have been used. The second term in the last formula can also be considered by using the quantum equation of motion in the form (\ref{gB}). Therefore, we need to shift only the mass parameter as follows
\begin{equation}
m^2\longrightarrow m^2+\frac{g^4}{12(4\pi)^4} L.
\end{equation}

\subsection{n=4 dimensional case}
\label{r4}
The divergencies in the effective action in the four-dimensional case follow from the equalities (\ref{Gd4}) and (\ref{gr4}), and formulae (\ref{4a1}) and (\ref{4a2}). Thus the contributions from the first two loops have the form
\begin{equation}\label{e38}
-\frac{g^2 L}{(4\pi)^2}\frac{(B,B)}{2}-
\frac{g^2 L}{2(4\pi)^2}\int_{\mathbb{R}^4}d^4x\,PS_4(x,x).
\end{equation}

In this case, we have only logarithmic divergencies. To renormalize the effective action, only the mass parameter should be shifted as follows
\begin{equation}\label{e39}
m^2\longrightarrow m^2-\frac{g^2}{(4\pi)^2} L.
\end{equation}

The four-dimensional case is super-renormalizable. Let  us see how the second singularity in  formula (\ref{e38}) can be cancelled.
Let $\sigma$ be a finite part of the $\ln\det(N)$ such that 
\begin{equation}
g\frac{\delta \sigma}{\delta v(x)}=\frac{\delta\sigma}{\delta B(x)}=-gPS_4(x,x),
\end{equation}
where $v(x)=-m^2+gB(x)$. When using the shift (\ref{e39}), the effective action $W[B]$ after the one-loop renormalization contains the term 
\begin{equation}\label{eqzv}
\left.\frac{1}{2}\left(\sigma+\frac{g^2L}{(4\pi)^2}\int_{\mathbb{R}^4}d^4x\frac{\delta\sigma}{\delta v(x)}\right)\right|_{m^2\to m^2+\frac{g^2}{(4\pi)^2}\ln L}.
\end{equation} 

However, all objects are constructed by using the Green's function. This means that they are functions of the field $v(x)=-m^2+gB(x)$.
At the same time, the operator 
\begin{equation}
\exp{\left(\frac{g L}{(4\pi)^2}\int_{\mathbb{R}^4}d^4x\,\frac{\delta}{\delta v(x)}\right)}
\end{equation} does a shift of the form
\begin{equation}
v(x)\longrightarrow v(x)+\frac{g^2L}{(4\pi)^2}.
\end{equation} 

 So one can see that formula (\ref{eqzv}) is equal to $\frac{1}{2}\sigma$ plus term, which is cancelled by the next high loop corrections. It is supposed that the same calculations can be done for a finite part of the two-loop correction using the high loop contributions.

\textbf{Remark 1.} Let us go back to the case $n=6$, where we noted that the term
\begin{equation}\label{dop_1}
-\frac{5}{6}\frac{g^2L}{2(4\pi)^3}\int_{\mathbb{R}^6} d^6x\,v(x)PS_6(x,x)
\end{equation}
exists. By $\sigma$ we denote the part of $\ln\det(N)$ such that $\frac{\delta\sigma}{\delta v(x)}=-PS_6(x,x)$. By analogy with the case $n=4$, we see that the term (\ref{dop_1}) is a part of exponential operator, which transforms the potential in the $\sigma$ from $v$ to $v+\frac{5}{6}\frac{g^2L}{(4\pi)^3}v$. At the same time, after one-loop renormalization we have the shift
\begin{equation}
v(x)=-m^2+gB(x)\to -Z_mZ^{-1}m^2+Z_0Z^{-1}gB(x)=v-\frac{5}{6}\frac{g^2L}{(4\pi)^3}v+...
\end{equation}
This means that the shifts cancel each other. A similar procedure should work in the high loops.

\section{Appendix A} 
It is very well known (see \cite{8,vas1,vas2}) that  the heat kernel $K(x,y;\tau)$ can be represented as a series in powers of a proper time $\tau$, when $\tau\to+0$. The coefficients $a_k(x,y)$, $k\in\mathbb{N}$, of the expansion satisfy the problem
\begin{equation}
\begin{cases}
a_0(x,y)=1; \\
(k+(x-y)^\mu \partial_\mu)a_k(x,y)=(\partial_\mu\partial^\mu+v(x))a_{k-1}(x,y),\,\,k>0,
\end{cases}
\end{equation}
and are called Seeley--DeWitt coefficients. They play an important role in physics. In a particular case, they give an asymptotic expansion of the Green's function $G_n(x,y)$ when $x\sim y$. Let us introduce some notations
\begin{equation}
(x-y)^{\mu_1\ldots\mu_k}=(x-y)^{\mu_1}\ldots(x-y)^{\mu_k},\,\,\,\,\,\,
\partial_{\mu_1\ldots\mu_k}=\partial_{\mu_1}\ldots\partial_{\mu_k},
\end{equation}
where $k\in\mathbb{N}$ and $\mu_i\in\{1,\ldots,n\}$. So we can write down the expansions for $n=3,4,5,6$:
\begin{equation}\label{Gd3}
G_3(x,y)=\frac{1}{4\pi r}-\frac{r}{8\pi}a_1(x,y)+PS_3(x,y)+o(r);
\end{equation}
\begin{equation}\label{Gd4}
G_4(x,y)=\frac{1}{4\pi^2r^2}-\frac{\ln (r\mu)^2}{16\pi^2}a_1(x,y)+\frac{r^2\ln (r\mu)^2}{64\pi^2}a_2(x,y)+PS_4(x,y)+o(r^2\ln r^2);
\end{equation}
\begin{equation}\label{Gd5}
G_5(x,y)=\frac{1}{8\pi^2r^3}+\frac{1}{16\pi^2r}a_1(x,y)-\frac{r}{32\pi^2}a_2(x,y)+PS_5(x,y)+o(r);
\end{equation}
\begin{equation}\label{Gd6}
G_6(x,y)=\frac{1}{4\pi^3r^4}+\frac{1}{16\pi^3r^2}a_1(x,y)-\frac{\ln (r\mu)^2}{64\pi^3}a_2(x,y)+\frac{r^2\ln (r\mu)^2}{256\pi^3}a_3(x,y)+PS_6(x,y)+o(r^2\ln r^2),
\end{equation}
where $r=|x-y|$, $PS_k(x,y)$ for $k=3,4,5,6$ are regular parts that depend on $\mu$, although $G(x,y)$ does not (see \cite{9}). The first three coefficients have the form (from \cite{10,11,12}):
\begin{multline}
\label{SD1}
a_1(x,y)=v(y)+\frac{1}{2}(x-y)^\mu\partial_\mu v(y)+\frac{1}{6}(x-y)^{\mu\nu}\partial_{\mu\nu}v(y)+\\+\frac{1}{24}(x-y)^{\mu\nu\rho}\partial_{\mu\nu\rho}v(y)+\frac{1}{120}(x-y)^{\mu\nu\rho\sigma}\partial_{\mu\nu\rho\sigma}v(y)+o(r^4);
\end{multline}
\begin{multline}
\label{SD2}
a_2(x,y)=\frac{1}{6}\partial_{\mu\mu}v(y)+\frac{1}{2}v^2(y)+\frac{1}{12}(x-y)^\mu\partial_{\mu\nu\nu} v(y)+\frac{1}{2}v(y)(x-y)^\mu\partial_\mu v(y)+\\+\frac{1}{40}(x-y)^{\nu\rho}\partial_{\nu\rho\mu\mu}v(y)+\frac{1}{8}((x-y)^\mu\partial_\mu v(y))^2+\frac{1}{6}v(y)(x-y)^{\mu\nu}\partial_{\mu\nu}v(y)+o(r^2);
\end{multline}
\begin{equation}
\label{SD3}
a_3(y,y)=\frac{1}{60}\partial_{\mu\mu\nu\nu}v(y)+\frac{1}{6}v^3(y)+\frac{1}{12}\partial_\mu v(y)\partial_\mu v(y)+\frac{1}{6}v(y)\partial_{\mu\mu}v(y).
\end{equation}

At the same time after applying the operator $N(x)$ to the equality (\ref{Gd6}) and using the Green's function definition, we have the following equality for $n=6$
\begin{equation}
\label{6eq}
-\frac{a_3(y,y)}{16\pi^3}-v(y)PS_6(y,y)-\partial_\mu\partial^\mu PS_6(x,y)\bigg|_{x=y}=0.
\end{equation}

\section{Appendix B}
\subsection{n=4:} In the four-dimensional case we have only two singularities, which can be obtained by formulae (\ref{Gd4}) and (\ref{SD1})
\begin{equation}
\label{4a1}
3\int_V d^4yd^4x \left(\frac{1}{4\pi^2r^2}\right)^2\left(-\frac{\ln (r\mu)^2}{16\pi^2}v(y)\right)\stackrel{\mathrm{IR}}{=}-\frac{3S^3}{2^8\pi^6}\int_{\mathbb{R}^4} d^4y\,v(y) L^2,
\end{equation}
\begin{equation}
\label{4a2}
3\int_V d^4yd^4x \left( \frac{1}{4\pi^2r^2}\right)^2PS_4(x,y)\stackrel{\mathrm{IR}}{=}\frac{3S^3}{2^4\pi^4}\int_{\mathbb{R}^4} d^4y\,PS_4(y,y) L,
\end{equation}
where $V=\mathbb{R}^4\times\{z\in\mathbb{R}^4:|z-y|<1/\Lambda\}$.

\subsection{n=5:}
In the five dimensional case we have five terms with singularities, among which there are not only logarithmic. So, to obtain them, we need to use the expressions (\ref{Gd5}), (\ref{SD1}), (\ref{SD2}), and the equality of the form
\begin{equation}
\label{q}
\int_{\mathbb{R}^k} d^kx\,(x_j-y_j)^2f(r)=\frac{1}{k}\int_{\mathbb{R}^k}  d^kx\,r^2f(r),\,\,\,\,\,\,j\in\{1,\ldots,k\},
\end{equation}
where $k\in\mathbb{N}$, and $f$ is a smooth function, quite good decreasing at the infinity.

So we have the contributions, which are proportional to $\Lambda^2$, $\Lambda$, and $L$:
\begin{equation}
\label{5a1}
3\int_V d^5yd^5x \left(\frac{1}{8\pi^2r^3}\right)^2\frac{v(y)}{16\pi^2r}\stackrel{\mathrm{IR}}{=}\frac{3S^4}{2^{11}\pi^6}\int_{\mathbb{R}^5} d^5y\,v(y)\Lambda^2;
\end{equation}
\begin{equation}
\label{5a2}
3\int_V d^5yd^5x \left(\frac{1}{8\pi^2r^3}\right)^2PS_5(x,y)\stackrel{\mathrm{IR}}{=}\frac{3S^4}{2^6\pi^4}\int_{\mathbb{R}^5} d^5y\,PS_5(y,y)\Lambda;
\end{equation}
\begin{equation}
\label{5a3}
3\int_V d^5yd^5x\,\frac{1}{8\pi^2r^3}\left(\frac{v(y)}{16\pi^2r}\right)^2\stackrel{\mathrm{IR}}{=}\frac{3S^4}{2^{11}\pi^6}\int_{\mathbb{R}^5} d^5y\,v^2(y) L;
\end{equation}
\begin{equation}
\label{5a4}
3\int_V d^5yd^5x \left(\frac{1}{8\pi^2r^3}\right)^2\frac{(x-y)^{\mu\nu}\partial_{\mu\nu}v(y)}{6\cdot16\pi^2r}\stackrel{\mathrm{IR}}{=}\frac{S^4}{2^{11}5\pi^6}\int_{\mathbb{R}^5} d^5y\,\partial_{\mu\mu}v(y) L;
\end{equation}
\begin{equation}
\label{5a5}
3\int_V d^5yd^5x \left(\frac{1}{8\pi^2r^3}\right)^2\left(-\frac{r}{32\pi^2}a_2(x,y)\right)\stackrel{\mathrm{IR}}{=}-\frac{3S^4}{2^{12}\pi^6}\int_{\mathbb{R}^5} d^5y\left(\frac{1}{3}\partial_{\mu\mu}v(y)+v^2(y)\right) L,
\end{equation}
where $V=\mathbb{R}^5\times\{z\in\mathbb{R}^5:|z-y|<1/\Lambda\}$.

\subsection{n=6: }In the six-dimensional case we have 13 contributions with singularities, among which there are not only logarithmic.
To calculate the divergencies, we use  expressions (\ref{Gd6})--(\ref{SD3}), and (\ref{q}). Then we have:
\begin{equation}
\label{6a1}
3\int_V d^6yd^6x\left(\frac{1}{4\pi^3r^4}\right)^2\frac{v(y)}{16\pi^3r^2}\stackrel{\mathrm{IR}}{=}\frac{3S^5}{2^{10}\pi^9}\int_{\mathbb{R}^6} d^6y\,v(y)\Lambda^4;
\end{equation}
\begin{equation}
\label{6a2}
3\int_V d^6yd^6x\left(\frac{1}{4\pi^3r^4}\right)^2\frac{(x-y)^{\mu\nu}\partial_{\mu\nu}v(y)}{6\cdot 16\pi^3r^2}\stackrel{\mathrm{IR}}{=}\frac{S^5}{2^{11} 3 \pi^9}\int_{\mathbb{R}^6} d^6y\,\partial_{\mu\mu}v(y)\Lambda^2;
\end{equation}
\begin{equation}
\label{6a3}
3\int_V d^6yd^6x\left(\frac{1}{4\pi^3r^4}\right)^2PS_6(y,y)\stackrel{\mathrm{IR}}{=}\frac{3S^5}{2^5\pi^6}\int_{\mathbb{R}^6} d^6y\,PS_6(y,y)\Lambda^2;
\end{equation}

\begin{multline}
3\int_V d^6yd^6x\left(\frac{1}{4\pi^3r^4}\right)^2\frac{(x-y)^{\mu\nu\rho\sigma}\partial_{\mu\nu\rho\sigma}v(y)}{120\cdot 16\pi^3r^2}\stackrel{\mathrm{IR}}{=}\\ \stackrel{\mathrm{IR}}{=}\frac{1}{ 2^{15}5 \pi^6}\int_{\mathbb{R}^6} d^6y\left(\sum_{\mu=1}^6\partial_{\mu}^4v(y)+2\sum_{\mu,\nu=1, \mu\neq\nu}^6\partial_\mu^2\partial_\nu^2v(y)\right) L
\end{multline}

\begin{equation}
\label{6a4}
\int_V d^6yd^6x \left(\frac{v(y)}{16\pi^3r^2}\right)^3\stackrel{\mathrm{IR}}{=}\frac{S^5}{2^{12}\pi^9}\int_{\mathbb{R}^6} d^6y\,v^3(y) L;
\end{equation}
\begin{equation}
\label{6a5}
6\int_V d^6yd^6x\,\frac{1}{4\pi^3r^4}\frac{v(y)}{ 16\pi^3r^2}\frac{(x-y)^{\mu\nu}\partial_{\mu\nu}v(y)}{6\cdot 16\pi^3r^2}\stackrel{\mathrm{IR}}{=}\frac{S^5}{2^{11} 3\pi^9}\int_{\mathbb{R}^6} d^6y\,v(y)\partial_{\mu\mu}v(y) L;
\end{equation}
\begin{equation}
\label{6a6}
6\int_V d^6yd^6x\,\frac{1}{4\pi^3r^4}\frac{v(y)}{16\pi^3r^2}PS_6(x,y)\stackrel{\mathrm{IR}}{=}\frac{3S^5}{2^5\pi^6}\int_{\mathbb{R}^6} d^6y\,v(y)PS_6(y,y) L;
\end{equation}
\begin{equation}
\label{6a7}
3\int_V d^6yd^6x \left(\frac{1}{4\pi^3r^4}\right)^2\frac{1}{2}(x-y)^{\mu\nu}\partial_{\mu\nu}PS_6(x,y)\stackrel{\mathrm{IR}}{=}\frac{S^5}{2^6\pi^6}\int_{\mathbb{R}^6} d^6y\,\partial_{\mu\mu}PS_6(x,y)\bigg|_{x=y} L;
\end{equation}
\begin{equation}
\label{6a8}
3\int_V d^6yd^6x\,\frac{1}{4\pi^3r^4}\left(\frac{(x-y)^\mu\partial_\mu v(y)}{2\cdot 16\pi^3r^2}\right)^2\stackrel{\mathrm{IR}}{=}\frac{S^5}{2^{13}\pi^9}\int_{\mathbb{R}^6} d^6y\, \partial_{\mu}v(y)\partial_{\mu}v(y) L;
\end{equation}
\begin{multline}
\label{6a9}
3\int_V d^6yd^6x
\left( \frac{1}{4\pi^3r^4}\right)^2\left(-\frac{\ln (r\mu)^2}{64\pi^3}a_2(y,y)\right)\stackrel{\mathrm{IR}}{=}\\\stackrel{\mathrm{IR}}{=}-\frac{3S^5}{2^{9}\pi^9}\int_{\mathbb{R}^6} d^6y \left(\frac{1}{6}\partial_{\mu\mu}v(y)+\frac{1}{2}v^2(y)\right)\left(\frac{1}{4}\Lambda^2-\frac{1}{2}\Lambda^2  L\right);
\end{multline}
\begin{multline}
\label{6a10}
3\int_V d^6yd^6x\left( \frac{1}{4\pi^3r^4}\right)^2
\left(-\frac{\ln (r\mu)^2}{64\pi^3}\frac{1}{2}(x-y)^{\nu\rho}\partial_{\nu\rho}a_2(x,y)\right)\stackrel{\mathrm{IR}}{=}\\\stackrel{\mathrm{IR}}{=}-\frac{S^5}{2^{11}\pi^9}\int_{\mathbb{R}^6} d^6y\left(\frac{1}{40}\partial_{\nu\nu\mu\mu}v(y)+\frac{1}{8}\partial_\nu v(y)\partial_\nu v(y)+\frac{1}{6}v(y)\partial_{\nu\nu}v(y)\right) L^2;
\end{multline}
\begin{multline}
\label{6a11}
3\int_V d^6yd^6x\left( \frac{1}{4\pi^3r^4}\right)^2\left(\frac{r^2\ln (r\mu)^2}{256\pi^3}a_3(y,y)\right)\stackrel{\mathrm{IR}}{=}\\\stackrel{\mathrm{IR}}{=}\frac{3S^5}{2^{12}\pi^9}\int_{\mathbb{R}^6} d^6y\left(\frac{1}{60}\partial_{\nu\nu\mu\mu}v(y)+\frac{1}{6}v^3(y)+\frac{1}{12}\partial_\nu v(y)\partial_\nu v(y)+\frac{1}{6}v(y)\partial_{\nu\nu}v(y)\right) L^2;
\end{multline}
\begin{equation}
\label{6a12}
6\int_V d^6yd^6x\,\frac{1}{4\pi^3r^4}\frac{v(y)}{16\pi^3r^2}\left(-\frac{\ln (r\mu)^2}{64\pi^3}a_2(y,y)\right)\stackrel{\mathrm{IR}}{=}-\frac{3S^5}{2^{11}\pi^9}\int_{\mathbb{R}^6} d^6y\left(\frac{1}{6}v(y)\partial_{\mu\mu}v(y)+\frac{1}{2}v^3(y)\right)L^2,
\end{equation}
where $V=\mathbb{R}^6\times\{z\in\mathbb{R}^6:|z-y|<1/\Lambda\}$.

\end{document}